\documentclass[%
 aip,
 amsmath,amssymb,
 reprint,%
]{revtex4-1}

\usepackage{graphicx}
\usepackage{dcolumn}
\usepackage{bm}

\usepackage[utf8]{inputenc}
\usepackage[T1]{fontenc}
\usepackage{mathptmx}
\usepackage{etoolbox}

\makeatletter
\def\@email#1#2{%
 \endgroup
 \patchcmd{\titleblock@produce}
  {\frontmatter@RRAPformat}
  {\frontmatter@RRAPformat{\produce@RRAP{*#1\href{mailto:#2}{#2}}}\frontmatter@RRAPformat}
  {}{}
}%
\makeatother
\begin{document}

\preprint{AIP/123-QED}

\title{Measurement of the laser pulse phase velocity in plasma channel for DLA optimization}
\author{E.M. Starodubtseva}
 \email{starodubtceva.em19@physics.msu.ru}
 \affiliation{Faculty of Physics, Lomonosov Moscow State University, 119991, Moscow, Russia}
\author{I.N. Tsymbalov}
 \affiliation{Faculty of Physics, Lomonosov Moscow State University, 119991, Moscow, Russia}%
 \affiliation{Institute for Nuclear Research of Russian Academy of Sciences, 117312, Moscow, Russia}
 \author{D.A. Gorlova}
 \affiliation{Institute for Nuclear Research of Russian Academy of Sciences, 117312, Moscow, Russia}
\author{K.A. Ivanov}
 \affiliation{Faculty of Physics, Lomonosov Moscow State University, 119991, Moscow, Russia}%
 \affiliation{Lebedev Physical Institute of Russian Academy of Sciences, 119192, Moscow, Russia}
 \author{R.V. Volkov}
 \affiliation{Faculty of Physics, Lomonosov Moscow State University, 119991, Moscow, Russia}%
 \author{A.B. Savel'ev}
 \affiliation{Faculty of Physics, Lomonosov Moscow State University, 119991, Moscow, Russia}%
 \affiliation{Lebedev Physical Institute of Russian Academy of Sciences, 119192, Moscow, Russia}

\date{\today}

\begin{abstract}

We demonstrate a novel, direct method for measuring the phase velocity $v_{\phi}$ of an intense laser pulse within a plasma channel -- the crucial parameter that controls the resonance condition in direct laser
acceleration (DLA). The technique exploits the second harmonic (SH) radiation generated at the channel sheath -- a phenomenon previously observed in laser-wakefield acceleration experiments. The SH emission angle is governed by a phase-matching condition that directly depends on $v_{\phi}$. Experimental measurements performed using a 1 TW, 10 Hz Ti:Sa laser system yield phase velocities in the range $v_{\phi}=(1.010-1.030)c$ for plasma electron densities in the range $n_e=(0.01-0.06)n_{cr}$. The diagnostic is validated through quasi‑3D particle‑in‑cell (PIC) simulations that reproduce the experimental conditions. This work provides a way to optimize DLA schemes by enabling in‑situ measurement of the laser pulse phase velocity in plasma channels.
\end{abstract}

\maketitle

\section{Introduction}

Laser-plasma acceleration represents one of the most promising applications of ultra-intense lasers for high-energy charged particle beams generation\cite{malka2012laser}. Laser plasma accelerators are of great interest because they can sustain large electric fields of the order of 100 GV/m. 
 
Laser plasma electron acceleration mechanisms, such as Laser Wakefield Acceleration (LWFA \cite{sprangle1988laser}) and Direct Laser Acceleration (DLA \cite{pukhov1999particle}), are achieved by forming laser plasma structures that influence the acceleration of electrons. The characteristic feature is the appearance of a positive charge volume, due to electrons being pushed from the axis by the laser ponderomotive force \cite{krushelnick1997plasma} or a specially pre-created due to capillary discharge \cite{kaganovich1999high}. This volume can be spherical ("bubble" in LWFA blow-out regime \cite{krushelnick2010laser}) or extended ("plasma channel", implemented in both DLA \cite{gahn1999multi, tsymbalov2019well} and LWFA mechanisms \cite{geddes2004high, wagner1997electron}). The main parameter of the plasma channel or bubble is the electron density distribution $n_e(r)$ within it.

Electron energy gain dynamics in the DLA mechanism is determined by the ponderomotive phase \cite{tsakiris2000laser} -- relative phase of the laser radiation and the electron betatron oscillations in the plasma channel ($\Phi=\omega t-\int\omega_b d t-kz$, where $k=\omega/v_{\phi}$ -- wave vector, $\omega$ -- incident laser pulse frequency, $\omega_b=\omega_p/\sqrt{2\gamma}$ -- betatron frequency, $\omega_p=\sqrt{4\pi n_e e^2/m}$ -- plasma frequency, $\gamma$ -- Lorentz factor, i.e. electron energy). Efficient electron energy gain is observed under resonance condition $\frac{d\Phi}{dt}=\omega(1-v/v_{\phi})-\omega_b=0=func(\gamma, \omega, v_{\phi}, n_e)$, where $v$ -- electron velocity, $\gamma$ is an electron parameter, $\omega$ is a fixed laser characteristic, $n_e$ is an initial electron density in plasma target, so only the phase velocity $v_{\phi}$ determines the resonance condition. 
Thus, DLA depends on the matching conditions of these phases. And it has a significant influence on the efficiency of electron acceleration \cite{starodubtseva2023low}. 

Following the above, the question of measuring phase velocity $v_{\phi}$ directly from the experiment is fundamentally important. The only diagnostic method that can be attributed to such a measurement of plasma channel/bubble parameters is the spectroscopy and the transverse interferometry methods of a discharge capillary plasma channel density measurement \cite{jang2011density, daniels2015plasma}. These methods are well-established for investigating large-scale structures characteristic of LWFA, where the typical channel/bubble diameter can be tens of micrometers \cite{krushelnick2010laser}. However, they are not applicable to the diagnosing narrow plasma channels in the DLA regime, whose transverse size is comparable to the laser wavelength and often does not exceed a few micrometers \cite{tsymbalov2019well, tsymbalov2020efficient}. This sub-wavelength scale makes it impossible to directly resolve the channel structure using classical interferometric methods. Additionally, DLA experiments usually utilize higher-density plasma \cite{rosmej2019interaction}, also making interferometry more challenging. 

The primary goal for DLA channel diagnostics is not merely to measure the electron density inside a plasma channel, but rather to determine the laser pulse phase velocity $v_{\phi}$ — the parameter that critically sets the resonance for electron acceleration. Consequently, the ideal diagnostics is one that is sensitive to $v_{\phi}$. 

This article presents a novel diagnostic technique for directly measuring the laser pulse phase velocity in a plasma channel. The method is based on the angular distribution of SH radiation generated at the plasma channel sheath, a phenomenon previosly observed in LWFA experiments \cite{gordon2008electro, helle2010measurement}. SH generation intrinsically determined by $v_{\phi}$ through a phase-matching condition. We demonstrate the technique experimentally using a TW-class laser system and validate it through comprehensive Particle-In-Cell (PIC) simulations, confirming its utility for optimizing DLA mechanism.

\section{Theoretical Background: Second Harmonic Generation from a Plasma Channel Sheath}

\begin{figure}[h!]
    \centering
    \includegraphics[width=1.0\linewidth]{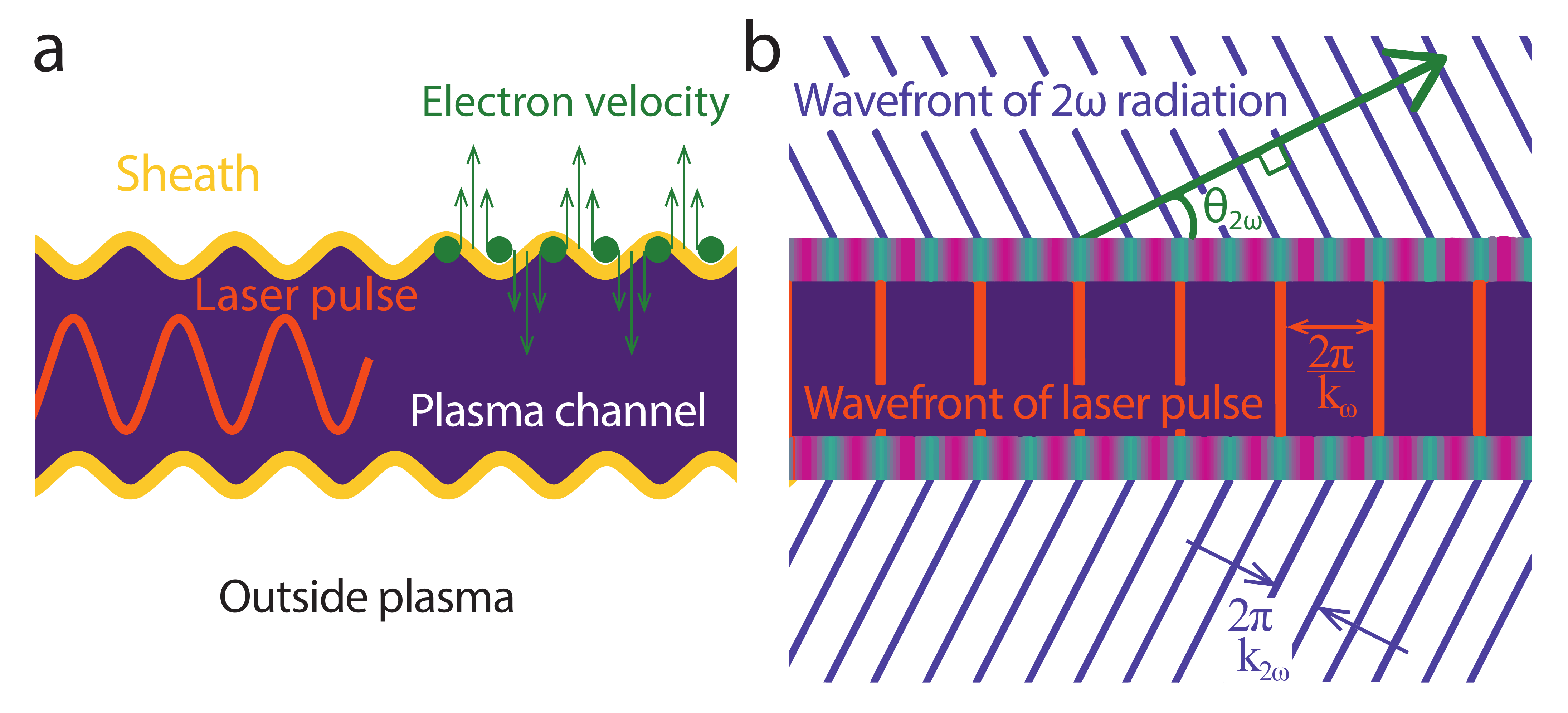}
    \caption{Mechanism of SH generation (a). Phase-matching conditions for SH generation from plasma channel sheath, ange of SH emission (see Eq. \eqref{eq1}) (b).}
    \label{fig01}
\end{figure}

When an intense laser pulse propagates through an inhomogeneous plasma, it drives electron density fluctuations in the plasma channel sheath \cite{kruer2019physics}. From the Gauss’s law, for a plasma with a dielectric permittivity $\epsilon(y)$ that varies transversely, one obtains $\epsilon(y)4\pi \rho+E_y \frac{\partial\epsilon(y)}{\partial y}=0$, where $\rho$ is the electron density in the sheath. The laser pulse electric field $E_y$ induces a density perturbation that oscillates in phase with the $E_y$: 
\begin{equation}
    \rho \sim -E_y\frac{\partial\epsilon(y)}{\partial y} \sim \cos(\omega t - kz)
    \label{x}
\end{equation}
(see yellow plasma channel sheath  in fig. \ref{fig01}a). From the Newton's equation of motion $m\frac{d\boldsymbol{v}}{dt}=-e\boldsymbol{E}$ the electron velocity $\boldsymbol{v}$ oscillations are phase-shifted by $ \pi/2$: $\boldsymbol{v} \sim -\sin(\omega t - kz)$ (see green arrows in fig. \ref{fig01}a). The resulting current density $\boldsymbol{j} = \rho \boldsymbol{v}$ is localized at the plasma channel sheath and represents a source of radiation at frequency $2\omega$ (see violet-sean colormap in fig. \ref{fig01}b).

The phase‑matching condition, derived from the geometry of the SH radiation emitting sheath (see fig. \ref{fig01}b, phase continuity of the source and SH radiation wavefronts must be ensured at the plasma channel sheath) leads to a simple relation for the emission angle $\Theta_{2\omega}$ measured from the channel axis \cite{gordon2008electro}:
\begin{equation}
\label{eq1}
    \cos{\Theta_{2\omega}} = \frac{2k_{\omega}}{k_{2\omega}}
\end{equation}
where $k_{2\omega} = \frac{2\omega}{c}\eta_{2\omega}$ is the wave number of the SH in a homogeneous medium with a refractive index $\eta_{2\omega}$. For the main harmonic, the plasma is inhomogeneous, and its wave number $k_{\omega}$ is defined by the phase velocity $v_{\phi}$: $k_{\omega}=\frac{\omega}{v_{\phi}}$.


The refractive index $\eta_{2\omega}$ can be expressed in terms of the electron density of the background plasma, $n_{e}$, as $\eta_{2\omega}=\sqrt{1-n_{e}/n^{2\omega}_{cr}}$, where $n^{2\omega}_{cr}=4n_{cr}$ is the critical density for the second harmonic. Substituting into Eq.\ref{eq1} and solving for $v_{\phi}$ yields the formula:
\begin{equation}
\label{eq2}
    v_{\phi}/c = \left(\cos{\Theta_{2\omega}}\sqrt{1-\frac{n_{e}}{4n_{cr}}}\right)^{-1}
\end{equation}

Thus, a measurement of the SH emission angle $\Theta_{2\omega}$, together with an estimate of the outside electron density $n_{e}$, provides a direct assessment of the laser phase velocity $v_{\phi}$ inside the channel.

\section{Experimental Setup}

\begin{figure}[h!]
\centering
\includegraphics[scale=0.23]{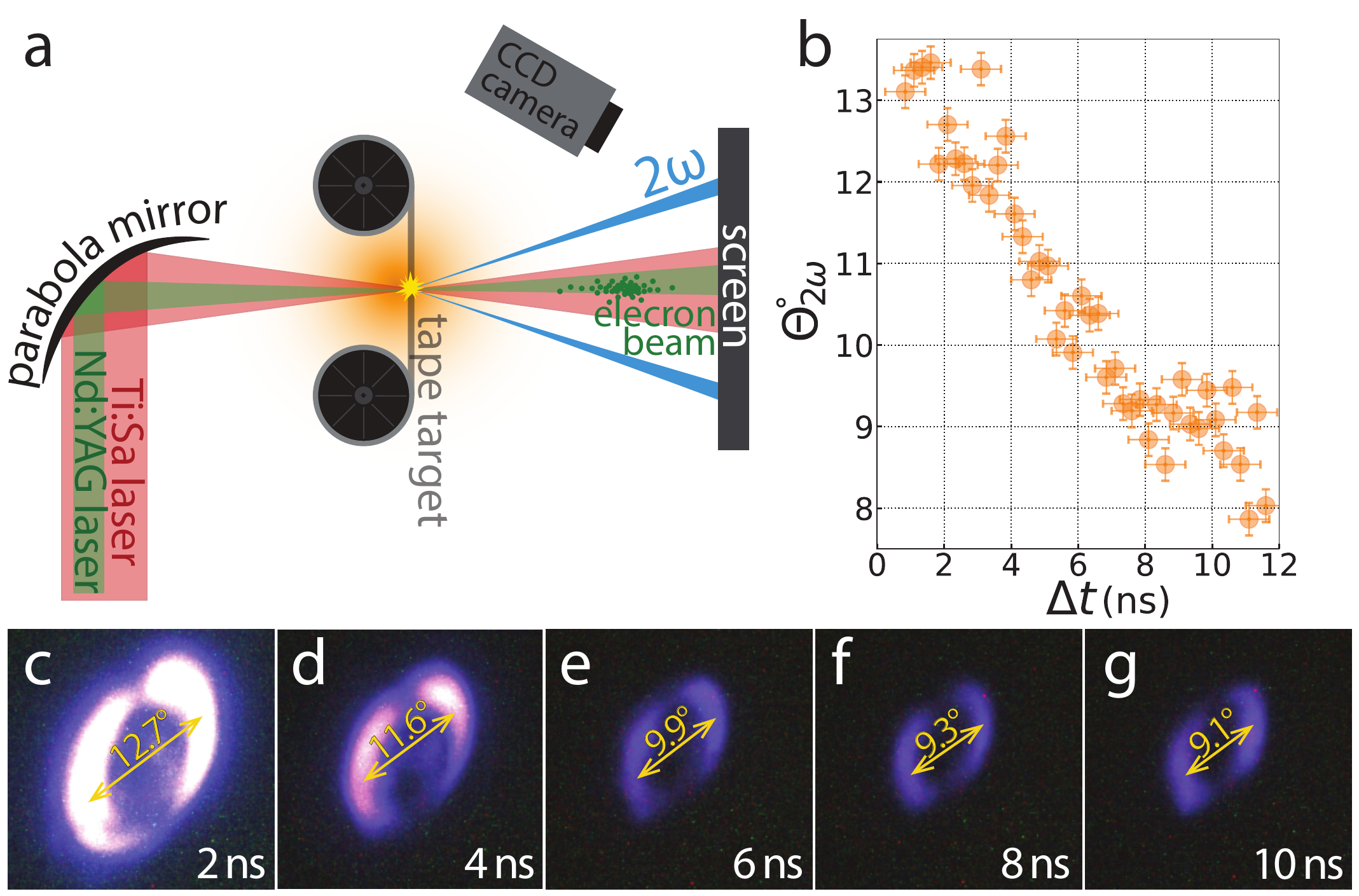}
\caption{\label{fig2} (a) Experimental setup. (b) The SH generation angle $\Theta_{2\omega}$ for different delay between ns and fs laser pulses $\Delta t$. (c-g) SH image from CCD camera for different $\Delta t$.}
\end{figure}

The experimental setup is demonstrated in fig. \ref{fig2}a. The experiment was performed using a 1 TW Ti:Sa laser system operating at a central wavelength of $\lambda=800$ nm. The laser delivered pulses with a duration of  50 fs (FWHM), linear polarization, and an energy of 50 mJ at a repetition rate of 10 Hz. The target consisted of a 12 $\mu$m thick PET (polyethylene terephthalate) tape.

To create a reproducible, expanding plasma plume, a separate Nd:YAG laser (1064 nm, 10 ns pulse duration, 200 mJ energy, 10 Hz) was used to ablate the front side of the film. By varying the temporal delay $\Delta t$ between the nanosecond (ns) ablation pulse and the main femtosecond (fs) pulse, the electron density $n_e$ at the interaction region could be controlled.

At a distance of 6.2 cm from the area of interaction, a screen on which the SH emission is observed was located. The screen image was recorded by CCD camera. A narrow band‑pass filter ($400\pm20$ nm) placed in front of the camera rejected plasma self‑emission and fundamental light. Examples of CCD images are shown in fig. \ref{fig2}c-g. The SH angle $\Theta_{2\omega}$ dependence on $\Delta t$ was obtained from the recorded images by measuring the radius of the emission ring (see fig. \ref{fig2}c-g) and calibrating with the known geometry of the experiment (see fig. \ref{fig2}b). The images recorded by the CCD camera are subject to distortion due to the camera's oblique viewing angle. The arrows in fig. \ref{fig2}c-g denote the horizontal direction, confirming that the SH angles measurement is reliable despite the distortion.

\section{PIC-simulation}

To validate the diagnostics and gain insight into the underlying physics, a series of quasi‑3D (azimuthally symmetric) PIC-simulations were performed using the SMILEI code \cite{derouillat2018smilei}. The simulations replicated the experimental parameters: a laser pulse with $a_0=1.5$, central wavelength of $\lambda = 800$ nm, a duration of $\tau=50$ fs (FWHM), focused to a spot size of $4\lambda$ (FWHM) in a simulation box of $100\lambda \times 40\lambda$. The spatial and temporal steps were $\frac{\lambda}{32}$ (longitudinal), $\frac{\lambda}{8}$ (transverse), and $\frac{\lambda}{48c}$, respectively.

\begin{figure}[h!]
\includegraphics[width=1.0\linewidth]{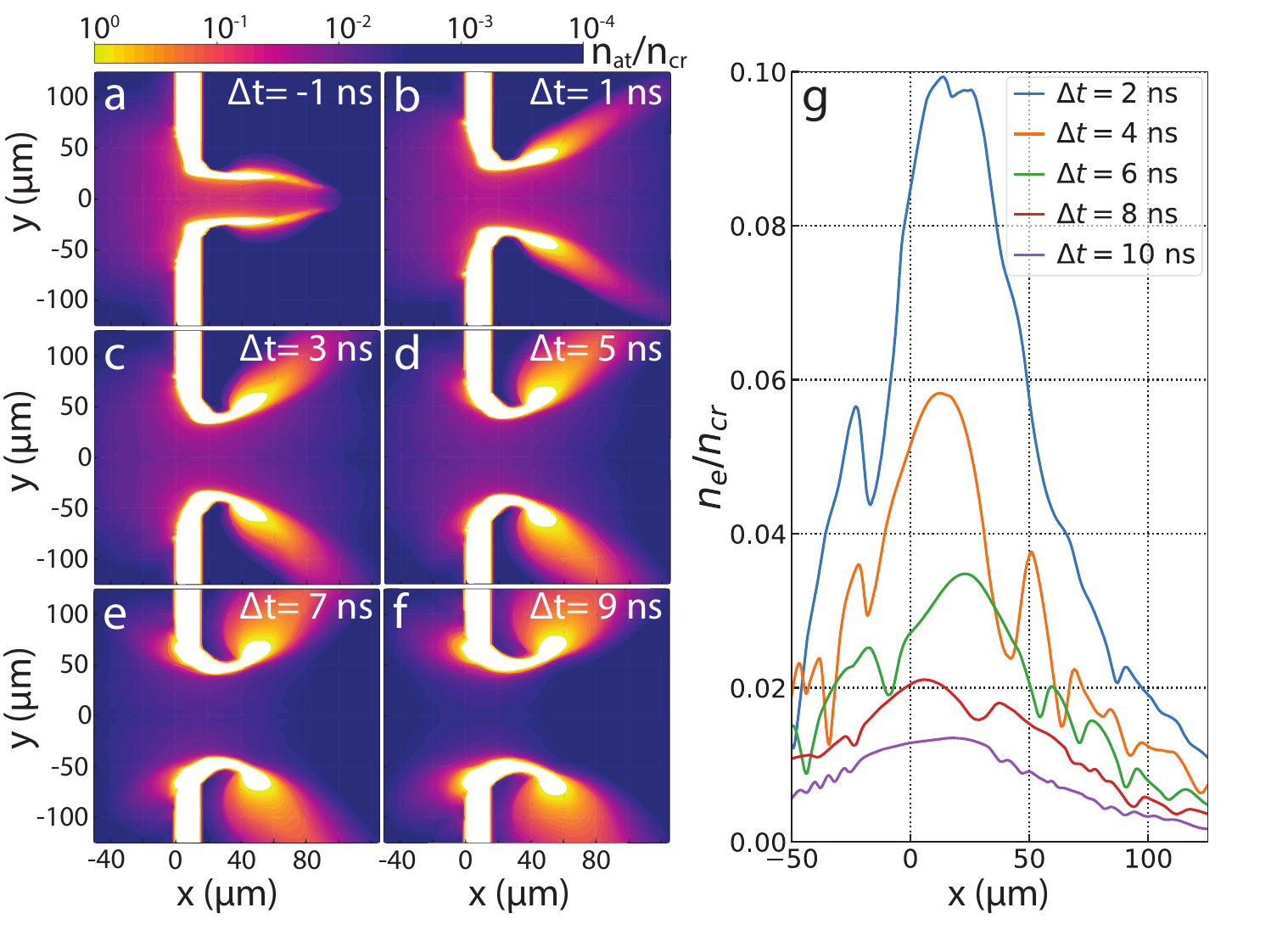}
\centering
\caption{ Hydrodynamic simulation of the PET tape ablation. Carbon atomic density $n_{at}$ in $n_{cr}$ for different delay between ns and fs laser pulses: $\Delta t=-1$ ns (a), $\Delta t=1$ ns (b), $\Delta t=3$ ns (c), $\Delta t=5$ ns (d), $\Delta t=7$ ns (e), $\Delta t=9$ ns (f). The electron density of the ablated film on axis $y=0$ for various values $\Delta t$ assuming that 4 electron ionized (g).}
\label{Hydro}
\end{figure}

The film target, consisting of neutral carbon atoms, was initialized with a density profile of the pre-plasma formed by the ns pulse, obtained from hydrodynamic simulations performed with hydrodynamic software package 3DLINE \cite{krukovskiy20173d} (see fig. \ref{Hydro}). In the PIC-simulations, the film density profile was approximated as carbon for computational simplicity. The carbon atomic density $n_{at}$ was rescaled to ensure that the electron density following the ionization of 4 carbon electrons was equivalent to the electron density of the ionized outer shells of PET atoms . The delay $\Delta t$ between the ns pre‑pulse and the main fs pulse was varied, corresponding to different stages of plasma expansion and hence different on‑axis densities. The validity of this approach is supported by the excellent agreement between PIC-simulation results and measured accelerated electron beam parameters from a similar experiment, as demonstrated in our previous study \cite{ivanov2024laser}. Thus, the successful use of hydrodynamic simulation results suggests that using the hydrodynamic density profiles as input data for PIC-simulations provides accurate estimates for this research. Field ionization was treated with the Ammosov‑Delone‑Krainov (ADK) model \cite{ammosov1986tunnel}.

\section{Results and Discussion}

\begin{figure*}
\includegraphics[width=1.0\linewidth]{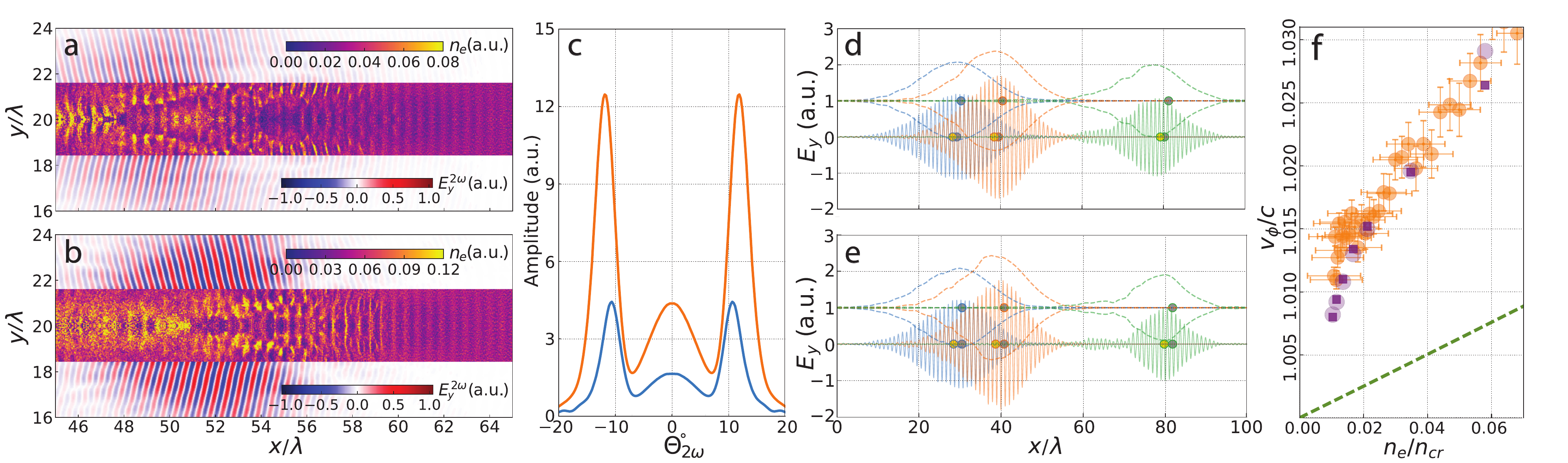}
\centering
\caption{PIC-simulation results. Electron density $n_e$ and Fourie filtrated SH electric field $E_y^{2\omega}$ $n_e=0.035n_{cr}$ (a), $n_e=0.058n_{cr}$ (b). SH angle distribution obtained from spatial Fourier transform of $E_y^{2\omega}$ for $n_e=0.035n_{cr}$ in blue (maximum at $\Theta_{2\omega}=10.7^{\circ}$) and $n_e=0.058n_{cr}$ in orange (maximum at $\Theta_{2\omega}=11.8^{\circ}$) (c). Laser pulse electric field $E_y$ (solid lines) on axis ($y=20\lambda$) at different time instants: $t=70\lambda/c$ (blue), $t=80\lambda/c$ (orange) and $t=120\lambda/c$ (green) and corresponding dashed pulse envelopes  at $y=21\lambda$ (approximately the plasma channel sheath) for  $n_e=0.035n_{cr}$ (d), $n_e=0.058n_{cr}$ (e). Phase velocity $v_{\phi}$ depending on electron plasma density $n_e$ in experiment (orange), PIC-simulation: calculated from SH angle $\Theta_{2\omega}$ by Eq. \eqref{eq2} (darkpurple squares) and "directly" measured (lightpurple circles) and the theoretical lower limit of the method's applicability in green (f).
\label{PIC SH}}
\end{figure*}

Figures \ref{PIC SH}a,b display characteristic electron density $n_e$ and temporal Fourier-filtrated electric field within the $(2\pm0.2)\omega$ band  $E_y^{2\omega}$ snapshots (at $t=98\lambda/c$) for different plasma densities $n_e=0.035n_{cr}$ (a), $n_e=0.058n_{cr}$ (b). The $E_y^{2\omega}$ in fig.\ref{PIC SH}a,b  clearly demonstrates the conical emission pattern of the SH. The corresponding SH angular distributions (see fig.\ref{PIC SH}c) were determined with spatial Fourier transform of $E_y^{2\omega}$, clear peaks are observed at positions determined by $\theta_{2\omega}=\arctan \frac{k_y}{k_x}$ ($\Theta_{2\omega}=10.7^{\circ}$ for $n_e=0.035n_{cr}$ and $\Theta_{2\omega}=11.8^{\circ}$ for $n_e=0.056n_{cr}$). 

As an independent phase velocity check, a “direct” measurement of the $v_{\phi}$ in the plasma channel was performed within the PIC-simulations. By tracking the longitudinal position of a specific zero‑crossing of the laser electric field $E_y=0$ and then linear fitting $x_{zero}(t)$ dependence, we obtained a precise value of $v_{\phi}$.

Examples of the laser pulse evolution within the plasma channel over time for $n_e=0.035n_{cr}$ and $n_e=0.058n_{cr}$ densities are shown in fig. \ref{PIC SH}d,e, correspondingly. The solid line represents the laser pulse on axis ($y=20\lambda$), while the dashed envelope corresponds to the laser pulse at $y=21\lambda$, which approximately coincides with the plasma channel sheath. The phase velocity of the laser pulse peak was measured by tracking a field zero-crossing located ahead of the pulse maximum (gray points in fig.\ref{PIC SH}d,e). For the presented cases, the obtained phase velocity values on the axis and at the sheath at the laser pulse peak are the following: $v_{\phi~axis}=1.0196\pm0.0002$, $v_{\phi~sheath}=1.0192\pm0.0001$ for $n_e=0.035n_{cr}$ and $v_{\phi~axis}=1.0291\pm0.0006$, $v_{\phi~sheath}=1.0297\pm0.0004$ for $n_e=0.058n_{cr}$. The errors represent the root-mean-square deviation of the linear fit to the $x_{zero}(t)$ dependence. The close agreement between the on-axis and sheath $v_{\phi}$ values indicates an essentially planar laser wavefront, implying that the phase velocity is uniform in the plane perpendicular to the channel axis.

The $v_{\phi}$ of several nearby laser pulse peak field zero-crossing on the plasma channel axis was also measured. As an example, the field zero-crossing located one wavelength away from the gray reference point is demonstrated (yellow points). For this zero-crossing, the measured phase velocities are $v_{\phi~axis}=1.0207\pm0.0005$ for $n_e=0.035n_{cr}$ and $v_{\phi~axis}=1.0308\pm0.0012$ for $n_e=0.058n_{cr}$. The close agreement of these three independently measured $v_{\phi}$ -- namely, (i) on the channel axis at the laser pulse peak, (ii) at the channel sheath at the pulse peak, and (iii) on the channel axis at locations shifted from the pulse peak -- strongly suggests that $v_{\phi}$ remains approximately constant both around the laser pulse peak and across the transverse extent of the plasma channel. This uniformity validates the applicability of the simplified model for SH generation used in our work.

The  Finite Difference Time Domain (FTDT) algorithm utilized in PIC-simulations inevitably exhibits numerical dispersion \cite{taflove2005computational}, which alters the phase velocity of electromagnetic waves in vacuum from the physical value c. In our simulation, we measured $v_{\phi_{2\omega}}/c = 0.9974$ for the SH in vacuum using the technique described above. To correctly apply the diagnostic formula \eqref{eq2} this numerical artifact must be accounted for. We introduced a corrected refractive index $\eta^{corr}_{2\omega}=\eta_{2\omega}+\Delta\eta$, where the correction $\Delta \eta = 1-c/v_{\phi_{2\omega}}$ compensates for the SH numerical dispersion.

In fig. \ref{PIC SH}f the phase velocity $v_{\phi}$ dependence on electron density $n_e$ is shown. Experimental results obtained from SH angle $\Theta_{2\omega}$ (see fig. \ref{fig2}b) by eq.\eqref{eq2} are marked in orange. The same PIC-simulation results calculated from the maxima of angular distribution (see fig. \ref{PIC SH}c) taking numerical dispersion into account are shown in darkpurple. The "directly" meashured $v_{\phi}$ in PIC-simulations are marked in lightpurple. Dependencies agreement confirms the applicability of this diagnostic method for measuring laser pulse phase velocity in the plasma channel.

Several factors can contribute to discrepancies between experiment and PIC-simulations. First, at higher electron plasma densities ($n_e\approx0.058n_{cr}$), the laser pulse undergoes significant self‑modulation (see laser pulse electric field evolution in fig. \ref{PIC SH}e compared to the low-density case $n_e=0.035n_{cr}$ in fig. \ref{PIC SH}d). This distorts the phase front and probably disrupts the SH emission, rendering the diagnostic invalid in that regime. We assume that this is the reason for the discrepancy in $v_{\phi}$ "directly" meashured and calculated from $\Theta_{2\omega}$ in PIC-simulation for $n_e=0.058n_{cr}$ (see lightpurple and darkpurple points in fig. \ref{PIC SH}f at $n_e=0.058n_{cr}$). However, the usable density range can be expanded by using higher intensities of laser radiation and a longer focusing scheme.

Second, accounting for the SH numerical dispersion as an additive to the refractive index may not be entirely correct. Although the coincidence of the calculated and measured $v_{\phi}$ suggests that this approach is applicable, at least in the specified range of $n_e$.

Third, jitter in the delay between the ns and fs pulses $\Delta t$, shot‑to‑shot laser energy fluctuations, and uncertainties in the absolute density calibration from hydrodynamic simulations all contribute to the discrepancies between experiment and PIC-simulations.



Despite these limitations, the overall agreement between the three independent determinations of $v_{\phi}$ -- from experimental SH angle, from SH angle in PIC, and "directly" measuring in PIC -- strongly supports the validity of the method.

A significant advantage of this method should be noted: the plasma channel does not need to be "ideal" for the diagnostic to work correctly. As seen in the fig. \ref{PIC SH}a,b, the channel has an uncertain structure. 
The term $\frac{\partial\epsilon(y)}{\partial y}$ in \eqref{x} does not have a periodic structure, so it cannot create new spatial Fourier components in $\rho$ and only broadens the existing ones. Consequently, the spatial Fourier components of $\rho$ repeat the spatial Fourier components of the laser electic field $E_y$. Therefore, the perturbations of the plasma channel induced by the ponderomotive force do not change the SH source spatial frequencies and so do not alter the SH emission angle $\Theta_{2\omega}$.

Many DLA studies, especially those focusing on efficient electron acceleration, are conducted under conditions of a significantly higher normalized vector potential ($a_0\gg 1$) and near-critical electron densities \cite{rosmej2019interaction, rosmej2020high}. However, in these regimes SH emission can be observed. According to SH generation theory, the limiting factor is reaching the angle of total internal reflection. The second harmonic will be observed at $v_{\phi}/c>1/\sqrt{1-\frac{n_e}{4n{cr}}}$. In fig. \ref{PIC SH}f, the line corresponding to this condition is marked in green. 

However, the diagnostics and interpretation of SH generation at these extreme parameters require accounting for several effects not affecting in lower-density and low-intensity regimes: 1) laser pulse self-modulation discussed above; 2) SH refraction due to density gradients at the plasma boundary (this effect is negligible at our low densities); and 3) at very high densities, corrections proposed by us for numerical dispersion in PIC-simulations may become insufficient or inaccurate.

\section{Conclusion}


We have presented a novel diagnostic technique for in-situ measurement of the phase velocity of an intense laser pulse inside a plasma channel. The proposed approach is based on recording the radiation angle of the second harmonic generated at the plasma channel sheath. The second harmonic emission phenomenon experiences the phase‑matching condition that makes the second harmonic angle depend on the laser pulse phase velocity.

Experimental implementation with a 1 TW laser system at a repetition rate of 10 Hz demonstrated the feasibility of the technique, yielding phase velocities in the range $v_{\phi}\approx(1.010-1.030)c$ for relevant plasma densities $n_{e_{out}}\approx(0.01-0.06) n_{cr}$. Comprehensive quasi‑3D PIC-simulations confirmed the underlying physics and provided independent validation through direct phase velocity measurement. 


Recent research consistently highlights that the achievable electron energy gain and beam charge are not governed by the laser intensity alone, but are critically dependent on the properties of the plasma channel formed during the interaction. As demonstrated by Hussein et al. (2021) \cite{hussein2021towards}, optimal DLA occurs at a specific plasma density ($\sim0.028 n_{cr}$), where a stable channel forms. Tang et al. (2024) \cite{tang2024influence} derived the optimal laser focal spot size for efficient energy gain, which depends on laser power and plasma density. Their experimental data show a clear peak in electron energy at a specific focal spot. Theoretical work by Arefiev et al. (2014) \cite{arefiev2014enhancement} demonstrates that substantial electron energy gain can occur due to a parametric instability of betatron oscillations, a condition determined by laser intensity and the electron density inside the plasma channel. Our previous work \cite{starodubtseva2023low} provides a theoretical framework showing that efficient injection and energy gain are determined by specific topological regimes in DLA phase space, which are directly dictated by plasma channel parameters -- the electron density and the laser pulse phase velocity inside the plasma channel.

Thus, several articles have demonstrated that DLA efficiency is strongly governed by plasma channel properties. Access to accurate, real-time information on these parameters is therefore critical for optimizing DLA performance. The proposed phase velocity diagnostics offers a straight way for electron acceleration optimization. Instead of relying on indirect methods, this diagnostic enables a direct, in-situ assessment of the resonance condition -- a requirement for sustained electron energy gain in the DLA mechanism.

\begin{acknowledgments}
This work was supported by RSCF grant №22-79-10087P with the use of equipment purchased under support of National project Science and Universities of Ministry of Science and Higher Education of the Russian Federation. E.M. acknowledges Foundation for the advancement of theoretical physics “BASIS” for the financial support.
\end{acknowledgments}

\section*{Data Availability Statement}
The data that support the findings of this study are available from the corresponding author upon reasonable request.

\nocite{*}
\bibliography{main}

\end{document}